\documentclass{wqcd03}                 

\usepackage{mathptmx}                 

\confname{QCD@Work 2003 - International Workshop on QCD, Conversano, Italy, 
14--18 June 2003}

\title{Lifetime measurement of atoms consisting of $\pi^{+}$ and $\pi^{-}$ mesons}

\author{V.~Yazkov\addressmark{a} \\ on behalf of DIRAC Collaboration}

\address[a]{Skobeltsyn Institute for Nuclear Physics of Moscow State Univeristy, Russia.}

\begin{document}

\begin{abstract}
The lifetime measurement of $\pi^{+}$ $\pi^{-}$ atoms ($A_{2{\pi}}$) with 10\%
precision provides in a model independent way the difference between the
S-wave $\pi\pi$ scattering lengths for isospin 0 and 2, $|a_0 -a_2|$, with 5\%
accuracy. The scattering lengths $a_{0}$ and $a_{2}$ have been calculated in 
Chiral Perturbation Theory (ChPT) with a precision better than 2.5\%. Therefore,
such a measurement will be a sensitive check of the understanding of chiral
symmetry breaking in QCD by giving an indication about the value of the quark
condensate, an order parameter of QCD. The method of observation and lifetime
measurement of $A_{2\pi}$ is discussed. Data on $A_{2\pi}$ production 
in Pt, Ni and Ti targets and preliminary results on the lifetime measurement 
are presented. 
\end{abstract}

\maketitle


\section*{Introduction}

Pionium or $A_{2\pi}$ is a hydrogen-like atom consisting of $\pi^+$ and 
$\pi^-$ mesons. The lifetime of this atom is inversely proportional to the 
squared difference between the S-wave $\pi\pi$ scattering lengths for isospin 
0 and 2, $|a_0 -a_2|$. This value is predicted by chiral perturbation theory 
(ChPT), and a measurement of the $\pi^+\pi^-$ atom lifetime provides 
a possibility to check predictions of ChPT in a model-independent way. 

\section{Present status of $\pi\pi$ scattering length investigation.}

In chiral perturbation theory (ChPT) the 2-loop calculation, exploiting Roy 
equations, leads to a precise prediction for $|a_0 -a_2|$ at the level of
1.5\%: $|a_0 -a_2|=0.265 \pm 0.004$~\cite{COLA00}. 

Experimental data on $\pi\pi$ scattering lengths can be obtained in an 
indirect way via the following processes:

\begin{enumerate}
\item $K_{e4}$ decay, $K^+ \Rightarrow \pi^+\pi^-e^+\nu_e$, using 
      a model-independent analysis. The most resent result obtained 
      from $4 \times 10^5$ $K_{e4}$ decays have been collected in 
      the experiment E865 (BNL). The analysis yields the $S$-wave 
      $\pi\pi$ scattering length $a_0= 0.216 \pm 0.013 ({\rm stat}) 
      \pm 0.002 ({\rm syst}) \pm 0.002 ({\rm theor})$ \cite{PISL03}. 
\item $\pi N \to \pi\pi N$ reaction. The latest measurement gives a value 
      $a_0 = 0.204 \pm 0.014 ({\rm stat}) \pm 0.008 ({\rm syst})$ 
      \cite{KERM98}. In this case the experimental procedure is not 
      strictly unambiguous, since the contribution of non-OPE components 
      ($\Delta$, $N^{\ast}$) can effect the reliability of the result 
      \cite{KERM98}. 
\end{enumerate}

\section{Theoretical motivation}

The $\pi^+\pi^-$ atom decays by strong interaction mainly into 
$\pi^0\pi^0$. The branching ratio of the alternative decay mode
$A_{2\pi} \to 2 \gamma$ is at the level of $4 \cdot 10^{-3}$. There 
is a relation \cite{URET61,BILE69} between $\Gamma_{2\pi^0}$ and $\pi\pi$ 
scattering lengths: $\Gamma_{2\pi^0} = C \cdot |a_0 -a_2|^2$ .

ChPT at next-to-leading order in isospin breaking provides correction 
to decay width $\Gamma^{NLO}_{2\pi^0} =\Gamma_{2\pi^0} ( 1+ \delta_{\Gamma}) , 
\delta_{\Gamma} = (5.8 \pm 1.2) \%$ . Using this correction the lifetime of 
$\pi^+\pi^-$ atoms is predicted to be~\cite{LJUB01}:

$$
\tau = (2.9 \pm 0.1) \cdot 10^{-15} s
$$

A measurement of the atom lifetime with an accuracy of 10\% allows 
to determine the corresponding $\pi\pi$ scattering length difference with 
5\% accuracy.

\section{Method of lifetime measurement}

The $A_{2\pi}$ are produced by Coulomb interaction in the final state 
of $\pi^+\pi^-$ pairs generated in proton--target interactions~\cite{NEME85,AFAN93}. 
After production $A_{2\pi}$ travel through the target and 
some of them are broken up due to their interaction with matter: ``atomic 
pairs'' are produced, characterized by small pair c.m. relative momenta 
$Q < 3$~MeV/$c$. These pairs are detected in the DIRAC setup. Other atoms 
annihilate into $\pi^0\pi^0$. The amount of broken up atoms $n_A$ depends 
on the lifetime which defines the decay rate. Therefore, the breakup 
probability is a function of the $A_{2\pi}$ lifetime.

\begin{figure}[htb]
\hbox to\hsize{\hss
\includegraphics[width=\hsize]{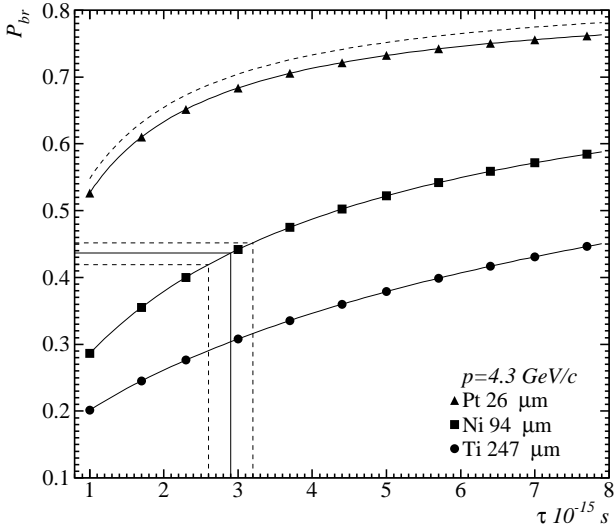}
\hss}
\caption{Dependence of the breakup probability $P_{\rm br}$ on $A_{2\pi}$ 
lifetime for three targets used in the DIRAC experiment: platinum of
26~$\mu$m, nickel of 94~$\mu$m and titanium of 247~$\mu$m thickness.}
\label{fig:break}
\end{figure}

The dependence of $P_{\rm br}$ on the lifetime $\tau$ is determined by 
the solution of differential transport equations~\cite{AFAN96}.
In Fig.~\ref{fig:break} the lifetime dependence of $P_{\rm br}$ is presented 
for three different targets used in the DIRAC experiment. The nickel target 
provides the best statistical accuracy for the same running time. 

Also $\pi^+\pi^-$ pairs from short-lived sources are generated in free state. 
Such pairs (``Coulomb pairs'') are affected by Coulomb interaction, too. 
The number of produced atoms ($N_A$) is proportional to the number of ``Coulomb 
pairs'' ($N^C$) with low relative momenta ($N_A=K \cdot N^C$). The coefficient 
$K$ is precisely calculable. And there are $\pi^+\pi^-$ pairs from long-lived 
sources ($\eta, K^0_s, \ldots$). Such pairs, not affected by final state 
interaction, are named ``non-Coulomb pairs''. 

\begin{figure}[htb]  
\hbox to\hsize{\hss
\includegraphics[width=\hsize]{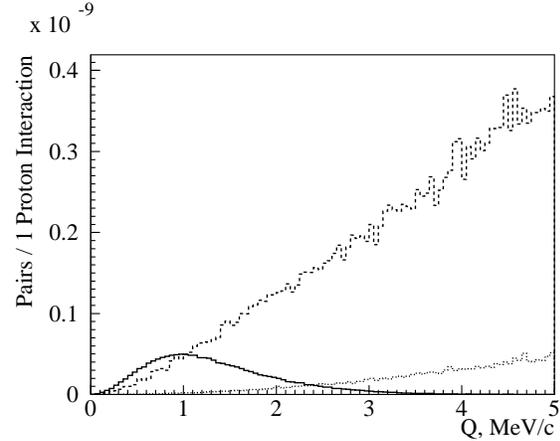}
\hss}
\caption{Simulated distributions of ``atomic pairs'' (solid line),
``Coulomb pairs'' (dashed line) and ``non-Coulomb pairs'' (dotted line).}
\label{fig:MC}
\end{figure}

In Fig.~\ref{fig:MC} simulated distributions of ``atomic pairs'', ``Coulomb'' 
and ``non-Coulomb'' pairs are presented. For relative momentum 
$Q > 3$~MeV/$c$ ``atomic pairs'' are practically absent, but in the region 
$Q < 2$~MeV/$c$ a fraction of them is essential, and this region provides 
the best effect-to-error ratio. 

The aim of DIRAC is to measure the $A_{2\pi}$ breakup probability 
$P_{\rm br}(\tau)$. $P_{\rm br}(\tau)$ is the ratio between the observed number 
of ``atomic pairs'' and the number of produced $\pi^+\pi^-$ atoms which is 
calculated from the measured number of ``Coulomb pairs''.

\section{Experimental setup}

\begin{figure}[htb]
\hbox to\hsize{\hss
\includegraphics[width=\hsize]{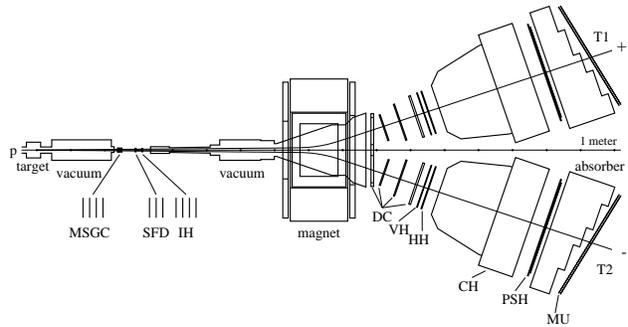}
\hss}
\caption{DIRAC setup. MSGC are microstrip gas chambers, SFD is 
a scintillating fiber detector and IH is a scintillation ionization 
hodoscope. Downstream the spectrometer magnet there are two identical 
arms T1 and T2. Each arm consists of drift chambers (DC), vertical (VH) 
and horizontal (HH) scintillation hodoscopes, threshold Cherenkov 
counters (CH), shower detectors (PSH) and scintillation muon detectors 
(MU)}
\label{fig:setup}
\end{figure}

The purpose of the DIRAC setup (Fig.~\ref{fig:setup}) is to detect $\pi^+\pi^-$ 
pairs with small relative momenta~\cite{ADEV03}. This setup is 
located at the CERN T8 beam area (East Hall). It became operational at 
the end of 1998 and uses the 24~GeV proton beam from PS accelerator. 

In Table~\ref{tab:setup} the main setup features are summarized. The setup 
resolution is shown for the projections of the relative c.m. momentum $Q$: 
$Q_L$ is a projection along the total pair momentum in the laboratory 
system, $Q_X$ and $Q_Y$ are the X- and Y- projections of the transverse 
component of $Q$. 

\begin{center}
\begin{table}[htb]
\caption{Setup features}
\label{tab:setup}
{\footnotesize
\begin{center}
\begin{tabular}{|p{4cm}|l|}
\hline
{} &{} \\[-1.5ex]
Angle of channel to proton beam & $5.7^{\circ}$ \\[1ex]
Channel aperture & $1.2 \cdot 10^{-3}$ sr \\[1ex]
Magnet bending power & 2.3 T$\cdot$m \\[1ex]
Momentum range & $1.2 \leq p_{\pi} \leq 7$ GeV/$c$ \\[1ex]
{} &{} \\[1ex]
Resolution on pair relative     & $\sigma_{Q_X}=\sigma_{Q_Y}=0.4$~MeV/$c$ \\[1ex]
momentum projections            & $\sigma_{Q_L}=0.6$~MeV/$c$ \\[1ex]
\hline
\end{tabular}
\end{center}
}
\vspace*{-13pt}
\end{table}
\end{center}

\section{Experimental data}

\begin{figure}[htb]  
\hbox to\hsize{\hss
\includegraphics[width=\hsize]{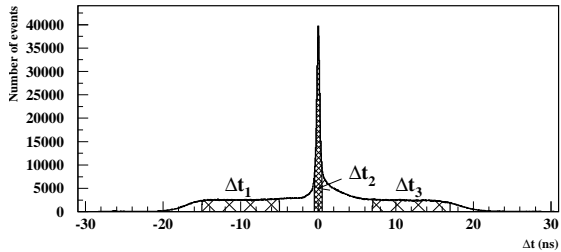}
\hss}
\caption{Distribution of events over time difference measured by 
vertical scintillation hodoscope (time of negative particle is 
subtracted from time of positive one). Regions $\Delta t_1$ and 
$\Delta t_3$ contain pure ``accidental pairs'' and $\Delta t_2$ one
contains a mixture of ``real'' and ``accidental'' pairs.}
\label{fig:dtime}
\end{figure}

The setup detects pairs of oppositely charged particles. Fig.~\ref{fig:dtime} 
shows the event distribution over time difference between positive and 
negative particles. The peak in the central part contains pairs of particles 
which are generated in the same proton--nucleus interaction (``real pairs'').
These pairs consist of ``atomic'', ``Coulomb'' and ``non-Coulomb'' pairs.
The wings of the distribution are formed by ``accidental pairs'' consisting 
of particles from different proton--nucleus interactions and so not 
affected by final state interaction. Therefore, the $Q$-distribution of 
``non-Coulomb''  and ``accidental'' pairs are very similar. Only a laboratory 
momentum distribution for ``accidental pairs'' is more hard than one for 
``real pairs'' due to energy conservation. It causes the difference for big 
values of pair relative momentum $Q$. But in DIRAC only events with small 
relative momenta are used. Admixture of ``accidental pairs'' under the central 
peak is subtracted by using the widths of the intervals $\Delta t_1$, 
$\Delta t_2$ and $\Delta t_3$. 

Eq.~\ref{eq:corf} shows the experimental correlation function $R(Q)$. This 
function consists of two parts. One part is a constant $N \cdot f$ as 
the result of dividing the distribution of ``non-Coulomb pairs'' 
($dN^{NC}_{\rm real}/dQ$) by the distribution of ``accidental pairs''. 
Here $N$ is a normalization parameter and $f$ the fraction of 
``non-Coulomb pairs''. 


\noindent
\begin{center}
\begin{equation}
R(Q)=\frac{\frac{dN_{\rm real}}{dQ}}{\frac{dN_{\rm acc}}{dQ}}=
\frac{\frac{dN^{NC}_{\rm real}}{dQ}+\frac{dN^{C}_{\rm real}}{dQ}}
{\frac{dN_{\rm acc}}{dQ}}=N \cdot [f + A_C(Q)]
\label{eq:corf}
\end{equation}
\end{center}

The second part of $R(Q)$ is $Q$-dependent: the number of ``Coulomb pairs'' 
($dN^{C}_{\rm real}/dQ$) with very small $Q$ is increased by Coulomb 
interaction. It should be emphasized that the Coulomb factor $A_C(Q)$ 
(Eq.~\ref{eq:corf}) is smeared compared to the theoretical one due to 
multiple scattering in the target and finite setup resolution.

\begin{figure}[htb]  
\hbox to\hsize{\hss
\includegraphics[width=\hsize]{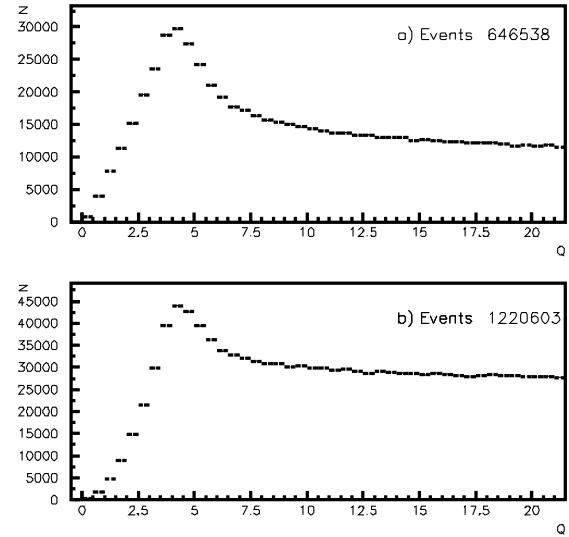}
\hss}
\caption{Experimental distributions of $\pi^+\pi^-$ pairs over relative
 momentum $Q$ for ``real'' pairs (a) and
 ``accidental pairs'' (b).}
\label{fig:Q}
\end{figure}

The experimental relative momentum distributions for ``real'' and 
``accidental'' pairs are shown in Fig.~\ref{fig:Q}. These data were collected 
with nickel target in 2001. 

\begin{figure}[htb]
\hbox to\hsize{\hss
\includegraphics[width=\hsize]{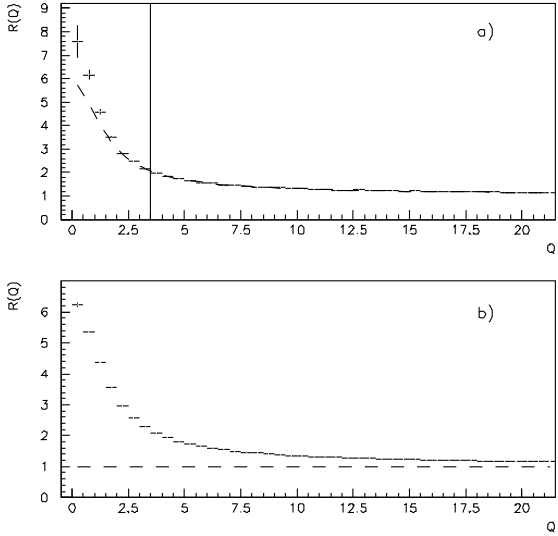}
\hss}
\caption{(a) The experimental correlation function of $\pi^+\pi^-$ pairs $R(Q)$ 
 (points with error bars) is fitted by an approximated correlation function 
 (smooth curve) in the region $Q > 3.5$~MeV/$c$. The excess for 
 $Q < 3.5$~MeV/$c$ occurs due to ``atomic pairs''. (b) The approximated 
 correlation function is the sum of correlation functions for ``Coulomb'' 
 (points) and ``non-Coulomb'' (dashed line) pairs.}
\label{fig:Qcor}
\end{figure}

Fig.~\ref{fig:Qcor} shows the experimental correlation function 
(Eq.~\ref{eq:corf}). In the region $Q > 3.5$~MeV/$c$ this function was fitted 
by an approximated correlation function which takes into account 
``non-Coulomb pairs'' (constant term) as well as ``Coulomb pairs'' 
($Q$-dependent term), the multiple scattering in the target and also 
the setup resolution. The fraction of ``non-Coulomb pairs'' $f$ is a free
parameter in the fit procedure (Eq.~\ref{eq:corf}). There is a good 
agreement between the experimental and approximated correlation functions for 
$Q > 2.5$~MeV/$c$. On the other hand in the region $Q < 2.5$~MeV/$c$ 
the experimental points are higher than the approximation function 
describing ``free pairs'' only. This excess is the signal of ``atomic pairs''. 

\begin{figure}[htb]  
\hbox to\hsize{\hss
\includegraphics[width=\hsize]{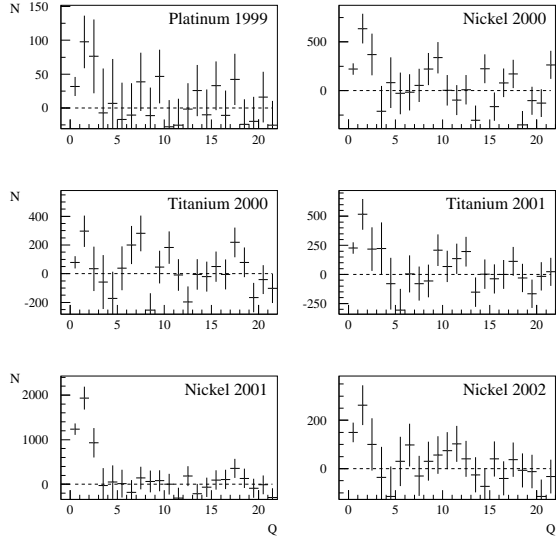}
\hss}
\caption{Signals of ``atomic pairs'' in the region $Q<3$~MeV/$c$ 
are obtained with different targets. For 2002 only data collected 
in 15 days are analyzed.}
\label{fig:eff}
\end{figure}

To calculate the number of ``atomic pairs'' the approximated correlation 
function has been transformed to a $Q$-distribution by multiplying it
with the distribution of ``accidental pairs'' (Eq.~\ref{eq:corf}). 
Differences of experimental and approximated distributions are shown in
Fig.~\ref{fig:eff} for data collected in 1999, 2000, 2001 and in 15 days of
2002.

Simulation shows that analysis of $Q_L$-distributions provides 
a decreasing of the systematic error by a factor $\sim 2$ and 
a small increasing (30\%) of the statistical error in comparison 
with analysis of $Q$-distributions. 

\begin{figure}[htb]  
\hbox to\hsize{\hss
\includegraphics[width=\hsize]{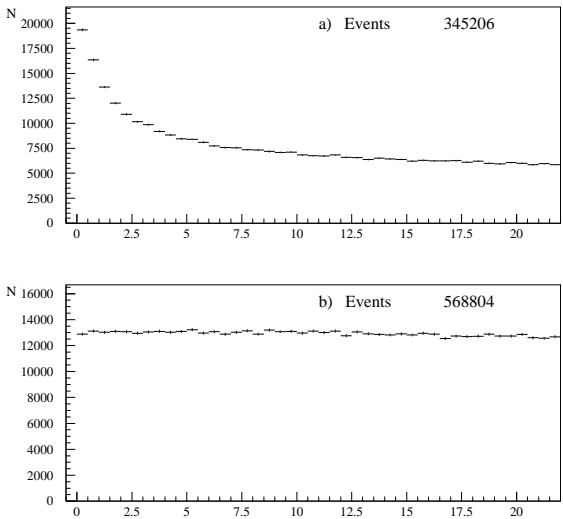}
\hss}
\caption{Experimental distributions of $\pi^+\pi^-$ pairs over 
$Q_L$-projection of pair relative momentum for mixture of ``real'' 
and ``accidental'' pairs (a) and ``accidental pairs'' (b).}
\label{fig:ni2001ra}
\end{figure}

The distribution over the absolute value of $Q_L$ for a mixture of ``real'' 
and ``accidental'' pairs and the distribution of ``accidental pairs'' only 
are shown in Fig.~\ref{fig:ni2001ra} for data collected with the nickel target 
in 2001. Data were selected with the criterion on transverse component of relative 
momentum $Q_T < 3$~MeV/$c$. Therefore a small value of $Q_L$ means a small value 
of $Q$. It is seen that the distribution of real pairs shows a peak at $Q_L=0$. 

\begin{figure}[htb]  
\hbox to\hsize{\hss
\includegraphics[width=\hsize]{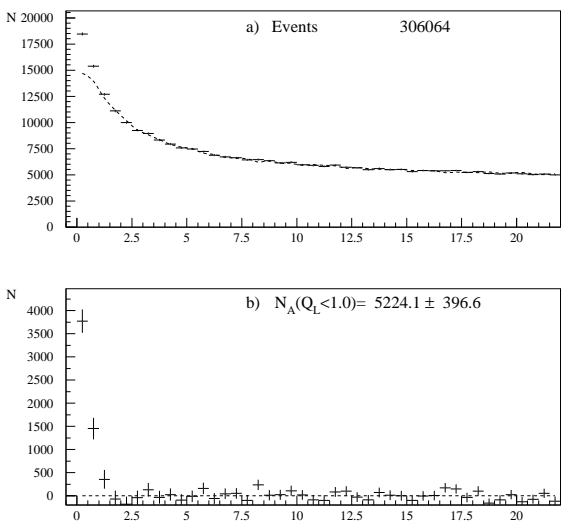}
\hss}
\caption{Experimental and approximated (dashed line) distributions of ``real''  
$\pi^+\pi^-$ pairs over $Q_L$ (a) and signal of ``atomic pairs'' in the region 
$Q_L<2$~MeV/$c$ is obtained using data collected in 2001 with a nickel target 
(b).}
\label{fig:ni2001re}
\end{figure}

The distribution of ``real pairs'' (Fig~\ref{fig:ni2001re}a) is fitted by the sum 
of approximated distributions of ``Coulomb'' and ``non-Coulomb'' pairs in 
the region $|Q_L| > 2$~MeV/$c$ where ``atomic pairs'' are absent. 
Approximated distributions are made on the base of ``accidental pairs'' 
applying a correlation function which takes into account the Coulomb interaction
in the final state, multiple scattering in the target, the resolution of the setup 
and difference of laboratory momentum spectra of ``real'' and ``accidental'' 
pairs. The fraction of ``non-Coulomb pairs'' and the normalization factor are 
free fit parameters. There is a good agreement between the experimental 
and approximated distributions for $Q_L > 2$~MeV/$c$. On the other hand in 
the region $Q_L < 2$~MeV/$c$ the experimental points are higher than 
the values of the approximation function describing ``free pairs'' only. 
This excess provides a number of ``atomic pairs'' and is shown in 
Fig~\ref{fig:ni2001re}b separately. The number of ``Coulomb pairs'' is 
calculated in the same fit procedure. 

The number of ``atomic pairs'' $n_A= 5224 \pm 397$ together with a number 
of ``Coulomb pairs'' and the theoretically known coefficient $K$ allows to
estimate a breakup probability value of $P_{\rm br}(\tau) =0.462$ with 
a statistical error 9.2\%. There is also a systematic error due to accuracy 
in description of multiple scattering in the target, membranes and detector planes 
(precision needed is 1\%) as well as in description of the detectors response. 

\begin{figure}[htb]
\hbox to\hsize{\hss
\includegraphics[width=\hsize]{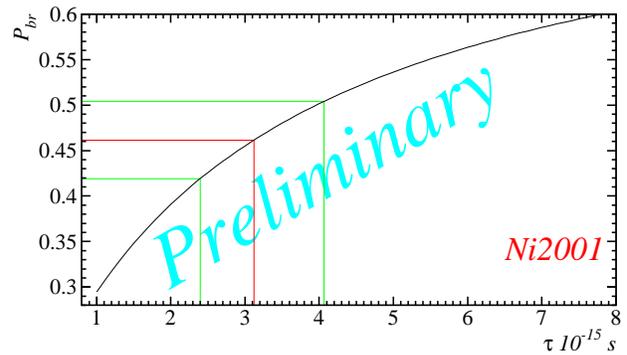}
\hss}
\caption{Preliminary estimation of $\pi^+\pi^-$ atom lifetime with
measured breakup probability.}
\label{fig:lifetime}
\end{figure}

Taking into account the dependence of breakup probability on lifetime 
(Fig.~\ref{fig:lifetime}) a preliminary estimation of the $\pi^+\pi^-$ atom 
lifetime yields 

$$
\tau = [3.12^{+0.92}_{-0.72} (stat) \pm 1. (syst)] \cdot 10^{-15} s
$$

\section*{Conclusions}

The DIRAC collaboration has built up a double arm spectrometer providing 
a resolution of 1~MeV/$c$ for low relative ($Q<30$~MeV/$c$) momentum 
of particle pairs and has successfully demonstrated its capability 
to detect $\pi^+\pi^-$ after 2 years of running time. 

In order to decrease systematic errors, dedicated measurements with 
a multi-layer nickel target and measurements of multiple scattering in all
detectors and setup elements have been performed at the end 2002 and during 
present run of 2003.

Preliminary results have been achieved by analyzing data collected in 2001. 
The statistical accuracy in the lifetime determination reaches 25\% and 
the systematic one is 30\%. The analysis of data collected in 2000 and 2002 
together with systematic error reduction allows us to improve 
the accuracy up to the level of 14\%.

\end{document}